# Real-Time Cognitive Evaluation of Online Learners through Automatically Generated Questions

Ritu Gala[1*], Revathi Vijayaraghavan[1], Valmik Nikam[1], Arvind Kiwelekar[2]

[1] Dept. of Computer Engg & IT, VJTI, Mumbai – 400019, India
[2] Dept. Of Computer Engg, Dr. Babasaheb Ambedkar Technological University, Lonere, Raigad – 402103, India
[*] Corresponding Author
```
rsgala_b17@ce.vjti.ac.in, rvijayaraghavan_b17@ce.vjti.ac.in,
       vbnikam@it.vjti.ac.in, awk@dbatu.ac.in
```

**Abstract.** With the increased adoption of E-learning platforms, keeping online learners engaged throughout a lesson is challenging. One approach to tackle this challenge is to probe learners periodically by asking questions. The paper presents an approach to generate questions from a given video lecture automatically. The generated questions are aimed to evaluate learners' lower-level cognitive abilities. The approach automatically extracts text from video lectures to generates wh-kinds of questions. When learners respond with an answer, the proposed approach further evaluates the response and provides feedback. Besides enhancing learner's engagement, this approach's main benefits are that it frees instructors from designing questions to check the comprehension of a topic. Thus, instructors can spend this time productively on other activities

## 1    Introduction

There has been a shift in recent times from traditional classroom learning towards remote learning. This implies that students can enrol in courses from different universities without needing to be physically present on the universities' campuses. Open online video course portals like MIT OpenCourseWare, Coursera, edX and others have gained much popularity because of their high-quality video lectures, which a student can access at any time and from anywhere.  To ensure that students have grasped the concepts explained in a  video lecture, there is a need to assess students on various topics presented in a  video lecture. Traditionally, a course instructor manually prepares a set of questions to evaluate the broad understanding of the material that has been taught. However, manually preparing questions is a time-consuming and cumbersome task. An instructor could use the time spent designing questions to prepare lecture material and presentations, which will be more fruitful and beneficial for the students and the course instructor himself.

The question designing task can be automated with advanced techniques from Natural Language Processing and Video Processing.  Automatic question generation is the task of generating questions from some kind of input. In our case, the input provided would be a video lecture for which assessment needs to be done. This paper presents the design of a system that generates questions from real-time video lecture. The objective behind generating questions is to increase online learners' engagement by periodically asking questions to check the comprehension of the topic being presented in a video lecture.

The rest of the paper is organised as follows: Section 2 discusses the background. The proposed methodology is elaborated upon in Section 3, and an analysis of our approach is presented in Section 4. Conclusion and future scope are presented in Section 5.

## 2  Related Work

Automatic Question Generation has observed great scope in the education domain. The research done in question generation has expanded exponentially in the past few years. Heilman[4] has developed a state of the art system that uses various Natural Language Processing(NLP) techniques and tools such as the Tregex expressions for T- Surgeon, BBN Identifier and Stanford Parser to generate questions from a given text. The majority of systems focus on automatic generation of questions from textual documents rather than the audio-visual medium. A few notable systems that work towards question generation from videos are specified as follows. Zhang et al.[10] have proposed the first model that generates questions of varying types by taking images or captions as input. Krishna et al.[3] present an automatic question generation system from MOOCs' videos at runtime. The transcripts of the videos use automatic discourse segmentation followed by content retrieval from Wikipedia documents to generate questions. The authors Y. Huang et al.[1] use a web scraper to crawl the transcripts of TED videos to generate two types of Multiple-Choice Questions (MCQs) that aim to assess listening comprehension, i. e., evaluating the listener's comprehension of the lecture's gist and the details described in it. The most related work to ours is Y. SKALBAN et al.[2] which produces questions from the subtitles of accompanying documentary videos. This is followed by referencing the timestamp of the subtitle and attaching a screenshot of the video along with the question.

Z. Zou et al. have conducted an extensive survey on the methods of object detection. Some of the widely used methods for object detection include the Regions with CNN (R-CNN), Faster R-CNN, Spatial Pyramid Pooling Networks (SPPNet), You Only Look Once (YOLO), Single Shot Multibox Detection (SSD), and RetinaNet [11]. Researchers have tested these methods on the PASCAL VOC Challenges [13,14], the MS-COCO dataset [15], the ILSVRC challenge [16], and the Open Images detection challenge [17]. Of these different methods, as shown in the results by J. Redmon et al. [12], the YOLO v3 is a faster and more accurate algorithm for real-time object detection and classification.

## 3 Proposed Methodology

The proposed method includes four stages as shown in Figure1. These are: (i) Video Processing, (ii) Question Generation, (iii) Object Detection, and (iv) Output the generated question.

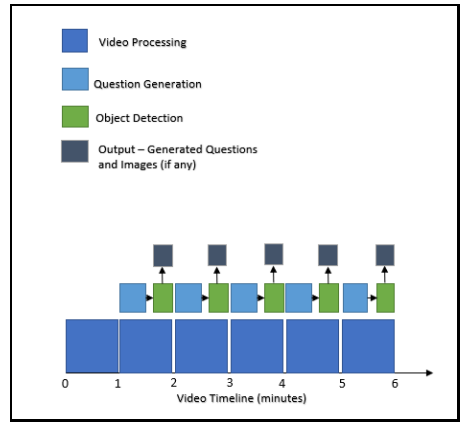

**Fig. 1.** Representation of timeline of the process

### 3.1 Video Processing

The first step in the proposed approach is video processing which is periodically carried out at 1-minute intervals. The objective of this step is to generate subtitles from the given video segment. The videos are split into 1-minute intervals for this purpose. The Google Speech-to-Text API can process audio streams of 1-minute duration synchronously. Otherwise, asynchronous processing of audios need to be done for audio streams that are greater than 1 minute in length. Also, synchronous processing is faster as compared to asynchronous processing. The extracted subtitles are stored in the '.srt' file format. First audio track is extracted from the video using the ffmpeg library. The input video format shall be in any of the formats supported by the ffmpeg library, for example, MPEG-1, MPEG-2, MPEG-4, ACM, etc. Once the audio track has been extracted, the audio track is passed to the Google Speech-to-Text API, converting the audio to text. Then the generated text is processed by the srt API to create subtitles in '.srt' format. The flow of this stage is shown in Fig. 2.

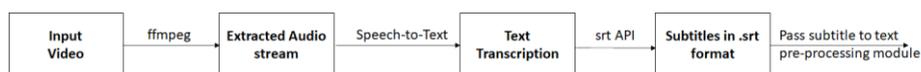

**Fig. 2.** The flow of Video Processing Stage

## 3.2 Question Generation

Once the video has been processed and subtitles have been generated, the subtitles are fed to the question generation module and the text preprocessing module. The subtitles first undergo the general NLP preprocessing. This involves tokenisation, stop words removal and stemming. The next step involved is diving the text into multiple subtopics. The subtopics are identified with the aid of the Text Tiling algorithm [7]. To identify subtopics, pseudo sentences are formed by grouping 20 consecutive stemmed words. The lexical cohesion between pseudo sentences is computed, which signifies the similarity between the sentences. These scores can then be plotted to show peaks and valleys. The depth score, which is the depth of the valley, is computed for each token-sequence pair. The depth scores are sorted and used to determine segment boundaries based on a threshold value. These computed boundaries are then stored for later processing.

Since the proposed system is being used for real-time videos and lectures, the kind of question that shall be generated must be comprehensive enough to test the content of the videos. At the same time, it should not be very complicated or difficult because students are still in the learning stage and they have yet to master the topic. With this in mind, the focus is on shallow question generation rather than deep question Generation. In other words, these questions shall evaluate the lower level cognitive abilities in terms of Bloom's Taxonomy i.e. Remember, Understand and Analyze [18]. For this purpose, we generate Wh-questions, as suggested by Heilman [4]. The preprocessing step's output is provided to the model to identify key phrases that could answer questions. The linguistic rules defined in [4] are applied to generate possible questions from a declarative statement. This is followed by removing the answer phrase from the sentence and inserting the question phrase. Finally, some basic post-processing is required to ensure proper formatting and punctuation.

### 3.3 Object Detection

To enhance the learning and testing experience, the question generated in the previous steps are linked with appropriate images wherever possible. To identify appropriate images from the video, an object detection algorithm called YOLO (You only look once) is used. First, we check if a question generated contains any of the class labels present in the training dataset used to build the custom YOLO model. For example, if the question generated is "What does this equation signify?" then we know that the question contains the class label "equation". Second, the source sentence which generated this question is located. Third, map the source sentence to a timestamp contained in the subtitles. Fourth, once the timestamp of the source has been found, run the fine-tuned YOLO object detection algorithm on the range of the sentence's timestamp to locate the object, which in this example would be the equation. Lastly, if the object is successfully found, then a screenshot of that time frame is attached to the question. If no object is found in the given time range, then the question is discarded as such a question would be considered a vague question. The flow of this stage is shown in Fig. 3, significant and the sample question with the image is shown in Fig. 4.

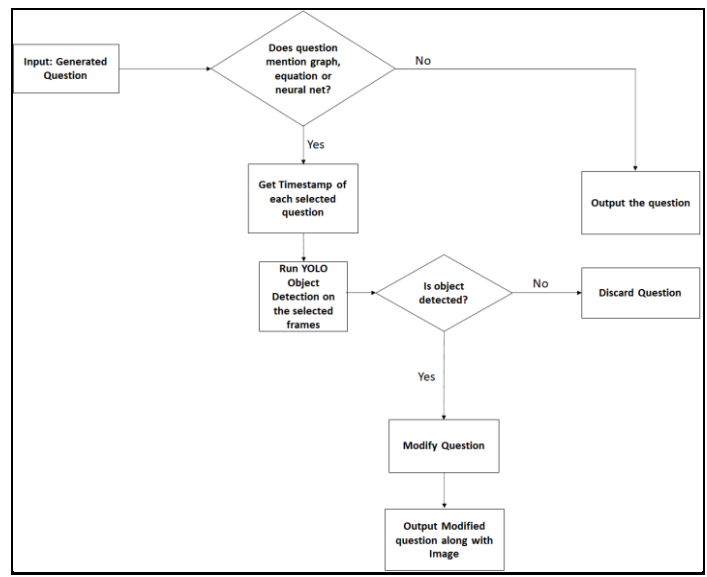

**Fig. 3.** Process for generating questions and linking images

**Fig. 4.** Sample image linked to the question [9]: "What does this equation signify?"

### 3.4    Output – Generated Question

This step performs the following three activities.

1. **Question Ranking**  The number of questions generated would be large, many questions of which might not be suitable for assessing the topic. To overcome this situation, the generated questions are ranked. For question ranking, the statistical method for ranking the questions described in [4] is used. The acceptability (y) of a question (x) is defined as

$$y = w^T f(x)$$

   Where f is a feature function that takes a question as input and returns a vector of real-valued numbers about different aspects of the question, and w is a vector of real-valued weight parameters for each feature in the feature vector returned by f. The features to be used for deciding a question's value would include the features described in [4] along with the feature of whether a question has an image linked to it or not.

2. **Assigning Questions**  The number of questions to be selected to assess a student depends on the video's total length. For instance, for a 40-minute long video, it is suggested to have 20 questions. This total number can be decided by the teacher beforehand. Each subtopic added previously would get proportionate weightage to the amount of time it was spoken about in the video (length of subtitles in total subtitles). Finally, we propose selecting random questions from the total for each student to allow unique learning.

3. **Feedback Design** Since the generated questions require students to type out their answers, qualitative feedback needs to be provided to each question. The qualitative feedback is generated by computing semantic similarity between the learner's answer and automatically extracted model answer**.** The use of Wordnet-based [5] similarity can be used to detect the correctness of the answer. Depending on the similarity score range, qualitative feedback value is assigned with either high similarity, medium similarity or low similarity. By providing real-time feedback, the student's learning is made more comprehensive.

## 4 Implementation Choices

This section explains the rationale behind the choices made while implementing the proposed system. These are:

1. **Topic Selection** To narrow the focus of our proposed system, we choose the topic of "*Machine Learning* (ML)" and "*Deep Learning (DL)*". Since there is an increased demand for these skills, the number of video lectures on them has consequently increased, giving us many videos on which our system can be applied. Further, these topics will help us decide the type of images used in our dataset to train the YOLO object-detection model.

2. **Preparing Dataset and Training** To deploy the YOLO model for our purposes, we have prepared our dataset by searching the web for images of equations, graphs and neural nets relevant to the Machine Learning and Deep Learning domain. To train our YOLO model, we used around 200 images of each class for accurate results. Once the model has been trained on our dataset, it is ready to be deployed for our application. Examples of relevant images included in the dataset are shown in Fig. 5, Fig. 6 and Fig. 7.

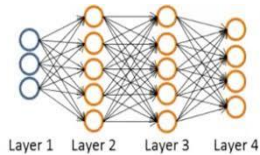

**Fig. 5.** Example of dataset image: Neural Network used in DL

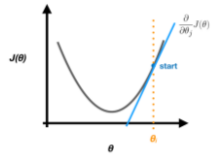

**Fig. 6.** Example of dataset image: Graph used in ML

$$J(\theta_0, \theta_1) = \frac{1}{2m} \sum_{i=1}^{m}(h_\theta(x^{(i)}) - y^{(i)})^2$$

**Fig. 7.** Example of dataset image: Equation used in ML

### 4.1 Evaluation of Approach

This section describes some of the metrics that can be used to evaluate proposed method's accuracy and efficacy, Evaluation can be performed in two ways i.e., intrinsic evaluation and extrinsic evaluation. The use of these two evaluation methods allow for a holistic appraisal of the questions generated.

**Intrinsic Evaluation.** The intrinsic evaluation aims to check the quality of questions generated and linking of questions to images. Firstly, the questions can be evaluated for how accurately the question generated from the subtitles themself, as well, how many semantic, syntactic errors and pedagogical errors present in it. The Heilman [4] model which is used for question generation serves as a base model for this purpose. Secondly, there needs to be an evaluation of the images linked to a question. This comprises two parts - evaluation for how well the fine-tuned YOLO model is, and whether the appropriate image has been linked for a given question. The fine-tuned

YOLO model is evaluated on the basis of the following metrics: mean average precision (mAP), for each of the categories of data that the model has been trained on. The number of epochs that the model is trained on can be adjusted, going up to 8000 epochs for best results. The mAP can be monitored for every 1000 epochs. The YOLO model giving the best mAP can then be selected and employed for the purposes of our image-linking. While monitoring the mAP for every 1000 epochs, we also monitor the loss to ensure that it is consistently decreasing. A mAP of approximately 95-99% is expected for the model to be extremely accurate. This will vary depending on the set of images used for training the YOLO model. Suitable image linking is gauged manually as we expect that the number of questions , wherein each question which has a linked image is checked to see if the image linked is (1) correct as per the timestamp and (2) relevant. Both of these evaluation metrics focus on questions which have images linked to them. These two accuracy based metrics would be defined as:

$$\text{correct\_timestamp\_accuracy} = \frac{\textit{number of questions with correct timestamp}}{\textit{total number of questions with images}}$$

$$\text{relevant\_accuracy} = \frac{\textit{number of questions which are relevant}}{\textit{total number of questions with images}}$$

A final metric that can be used is to check every question to see if an image should have been linked to a given question. However considering the large corpus of questions generated from each video, this manual task is out of scope for long videos.

**Extrinsic Evaluation.** Extrinsic Evaluation would focus on how much this post lecture assessment aided students' holistic learning. This can be evaluated using a survey, wherein after conducting a suitable number of lectures, students are asked to provide a rating (on a scale of 1 to 10) on whether they thought the questions were pertinent, appropriate and helpful for their learning.

## 5    Conclusion and Future Work

The paper describes a video processing based approach to automatically generate questions to evaluate cognitive abilities of online learners. The main activities performed for this purpose include: (i) extracting subtitles from the videos (ii)processing text to generate questions, (iii) linking questions to images and graphs, and (iv)Selecting appropriate questions by ranking their relevance to the content in the videos. The approach presented in this paper has numerous advantages such as timely generation of questions, improved testing experience by linking questions to images, adaptability of approach to other domains or disciplines, and comprehensive assessment of learners. A prototype of the approach has been implemented in a few video lectures from courses on Machine Learning. The approach needs to be evaluated against the various engagement and quality of questions generated. Also, the training dataset used in the YOLO model needs enhancement in terms of more images and graphs. During initial testing of the approach, we observed increased engagement in terms of time spent while watching the videos and increased comprehension of the topic presented in the video lectures.

## Author Biography

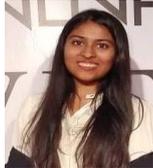 **Ritu Gala**, is pursuing her Bachelor of Technology in Computer Engineering at Department of Computer Engineering and Information Technology, VJTI Mumbai. Her research interests include Artificial Intelligence, Deep Learning, Machine Learning, and Blockchain. She has worked as a research intern for Citispotter Limited and Fourth Frontier, and is currently working as a student researcher in VJTI Blockchain and AI Lab. She has worked as a summer analyst and gained industry experience in Goldman Sachs. Her personal webpage is at https://www.linkedin.com/in/ritu-gala/

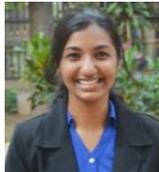 **Revathi Vijayaraghavan,** Student of Bachelor of Technology(Computer Engineering) at Dept. of Computer Engg and IT, VJTI Mumbai. Her research interests include Machine Learning, Deep Learning, Computer Vision and Blockchain. She has participated in several Kaggle competitions, and stood in the top 15 for the COVID-19 Global Forecasting. She has also worked as a student researcher in the VJTI Blockchain and AI Lab, and has worked in the industry as a Technology Analyst for Morgan Stanley. During her time in the Blockchain and AI Lab, she has co-authored 3 research papers. Her personal webpage is at  https://revathivijay.github.io/revathi-vijay/ .

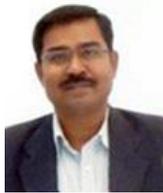 **Dr V.B. Nikam,** Associate Professor, Computer Engg & IT, VJTI Mumbai, has done Bachelor, Masters and PhD in Computer Engineering. He has 25yrs experience. He was felicitated with IBM TGMC-2010 DRONA award by IBM Academic initiatives. He is Senior Member (CSI), Senior Member (IEEE), and Senior Member (ACM). He worked on BARC ANUPAM Supercomputer. He was invited to JAPAN for K-Supercomputer study tour in 2013. He has received a grant-in-aid form NVIDIA for CUDA Teaching and Research, 2013. Presently, He is PI and Coordinator, Faculty Development Center (Geo-informatics, Spatial Computing and BigData Analytics) funded by MHRD, Govt of India. He works in the area of Data Mining and Data Warehousing, Machine Learning, Geoinformatics, Big Data Analytics, Geo-Spatial Analytics, Cloud Computing, GPU/High Performance Computing. You may visit www.drvbnikam.in for details.

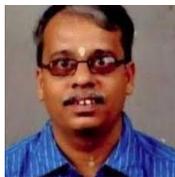 **Arvind Kiwelekar** is a Professor, Computer Engineering, at Dr Babasaheb Ambedkar Technological University Lonere. He completed his PhD from Indian Institute of Technology Bombay. His research interests include Software Architecture, Applied Artificial Intelligence and Ontology. He is actively engaged in conducting faculty development programs to train teachers on pedagogical theories, learning and teaching technologies, and providing training on emerging technologies such as Artificial Intelligence, Blockchain Technologies, Cognitive Modelling and Fog Computing.
You may visit awk-group.net for more details.